\documentclass[aps,pra,twocolumn,groupedaddress,showpacs]{revtex4}
\usepackage{graphicx}
\usepackage{hyperref}
\begin{document}
\title{Conditional preparation of a non-classical state in the continuous variable regime: theoretical study}
\author{J. Laurat}
\author{T. Coudreau} \email{coudreau@spectro.jussieu.fr}
\author{N. Treps}
\author{A. Ma\^{\i}tre} \author{C. Fabre } \affiliation{Laboratoire
Kastler Brossel, UPMC, Case 74, 4 Place Jussieu, 75252 Paris cedex
05, France}

\date{\today}

\begin{abstract}
We study the characteristics of the quantum state of light
produced by a conditional preparation protocol totally performed
in the continuous variable regime. It relies on conditional
measurements on quantum intensity correlated bright twin beams
emitted by a non-degenerate OPO above threshold. Analytical
expressions as well as computer simulations of the selected state
properties and preparation efficiency are developed and show that
a sub-Poissonian state can be produced by this technique.
Projection onto a given trigger value is studied and then extended
to a finite band. The continuous variable regime offers the unique
possibility to improve dramatically the preparation efficiency by
choosing multiple selection bands and thus to generate a great
number of sub-Poissonian states in parallel.
\end{abstract}

 \maketitle

\section{Introduction}
\label{sec:intro}Conditional state preparation opens the
possibility to generate a wide range of non-classical states of
light. Quantum correlation between two modes -a signal and a
trigger- is the prerequisite for such a general procedure.
Pre-assigned events on the trigger condition the readout of the
signal and cause a non-unitary state reduction.

Various protocols have been proposed and implemented successfully
in the photon counting regime. For instance, this strategy has
been widely used to generate single photon Fock state. The
required two-mode correlation can be obtained by an atomic cascade
\cite{Aspect} or by the more efficient technique of parametric
down conversion \cite{fock1}. The single-photon state is created
from the initial two-mode state when a photocount event is
recorded in the trigger path. Two output ports of a beam splitter
can also provide the required correlation: it has been shown
theoretically that a squeezed vacuum transmitted through a
beam-splitter is reduced to a Schr\"{o}dinger-cat-like state when
a photon is detected on the reflected port \cite{dakna}. Non
classical photon states have also been experimentally prepared by
triggering on a given atomic state \cite{atom}.

Recently, several conditional preparation protocols were
introduced for the continuous variable regime where a continuously
varying photocurrent is measured instead of single photon counts.
For instance in \cite{Schiller}, continuous measurements are
triggered by photocounts. The triggering condition and the
characterization of the selected state can also be done both in
the continuous variable regime. Some theoretical protocols have
been proposed relying on two-mode squeezed vacuum generated by an
optical parametric amplifier \cite{vacuum}.

The scheme we propose relies on quantum intensity correlated
bright twin beams -signal and idler- generated by a non-degenerate
optical parametric oscillator (ND-OPO) pumped above threshold
\cite{traditional}. The procedure, sketched in Fig. 1, is the
following : among all the recorded values of the signal beam
intensities, one only keeps the events occurring in the time
intervals when the idler intensity takes a given pre-assigned
value, or more precisely, lies around this value within a small
intensity range. All other time intervals are discarded. In the
present paper, we study theoretically this conditional preparation
technique. Its experimental implementation and the corresponding
results have been detailed in \cite{laurat}.

\begin{figure}
\includegraphics[width=.9\columnwidth]{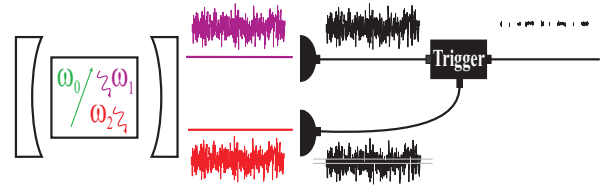}
\caption{\label{principe}Proposed conditional measurement protocol
in the continuous variable regime. A ND-OPO generates above
threshold orthogonally polarized and quantum intensity correlated
bright twin beams which are detected by high efficiency
photodiodes. The continuously varying signal photocurrent is kept
only when the idler photocurrent takes a given value or falls
inside a narrow band.}
\end{figure}
This paper is organized as follows. In section \ref{sec:opostate},
we describe the quantum state generated by a ND-OPO above
threshold. Conditioning on a given value of the idler intensity
results in a state reduction which generates a sub-Poissonian
state. This procedure, presented in section \ref{sec:onevalue}, is
then extended to the case of a selection interval of bandwidth
$\Delta$ around a given value in section \ref{sec:band}. The state
reduction is characterized according to the initial correlation,
individual noise on each beam and position and width of the
conditioning band. Section \ref{sec:simulation} proposes a
Monte-Carlo simulation of the reduced state properties. Section
\ref{sec:multiband} is then devoted to multiple-band selection, a
possibility to significantly increase the efficiency of the
conditional strategy which is only offered by schemes in the
continuous variable regime. In section \ref{sec:conclusion}, the
main conclusions of the paper are summarized and possible
extensions to more exotic non-classical states are discussed.

\section{Quantum state produced by a ND-OPO above threshold}
\label{sec:opostate} The starting point of the theoretical
analysis of the present conditional measurement is the precise
knowledge of the quantum state produced by a non-degenerate OPO,
from which the conditionally prepared state will be deduced by
state reduction. When the twin photons are produced by spontaneous
parametric down-conversion, the quantum state of the system is
well-known and given by \cite{milburn}
\begin{eqnarray}
|\Psi\rangle &=& e^{\left(\lambda \hat{a}_1^{+}\hat{a}_2^{+}
-\lambda^* \hat{a}_1\hat{a}_2\right)}|0,0\rangle \nonumber \\ &=&
(\cosh\lambda)^{-1}\sum_{n=0}^{\infty} (\tanh\lambda)^n
|n,n\rangle
\end{eqnarray}
where the indices 1 and 2 refer respectively to the signal and
idler modes and $\hat{a}_i$ (resp. $\hat{a}_i^{+}$) is the
annihilation (resp. creation) operator of a photon in mode $i$.
$\lambda$ is proportional to the pump amplitude and crystal
non-linearity. The state $|n_1,n_2\rangle$ is a Fock state with
$n_1$ photons in the signal mode and $n_2$ photons in the idler
mode. When the parametric down-conversion efficiency is very weak,
which is experimentally the most frequent case, the state of the
system reduces to the simple entangled state
\begin{eqnarray}
|\Psi\rangle \simeq |0,0\rangle + \lambda |1,1\rangle
\end{eqnarray}
When a single photon is detected in the idler mode, the system
collapses by the state reduction process into the signal mode
single photon state $|1\rangle$, as is well known.

To the best of our knowledge, there is so far no complete theory
giving the state of light produced by an OPO in the
Schr\"{o}dinger picture, comparable to the Lamb theory giving the
density matrix of the light produced by a laser \cite{lamb}. The
OPO has been described instead in most theoretical approaches in
the Heisenberg picture, in terms of operators, which is not useful
in the present case. Only Graham {\it et al.} \cite{graham} have
considered this problem, mainly to determine the phase diffusion
properties of the OPO, and not its joint photon-number
distribution.

In the absence of a complete theory, we will use here a
semi-phenomenological approach, relying on both theoretical
considerations and experimental observations
\cite{traditional,laurat} : OPOs produce above threshold
phase-coherent light in both signal and idler modes which have a
Gaussian statistics and present close to threshold a large excess
noise compared to the shot noise level. We will call $F$ the Fano
factor of this photon distribution defined as the noise variance
normalized to the one of a Poissonian distribution of same mean
power: close to threshold, this factor is large. Furthermore, one
knows that in the absence of losses and in the case of a single
output mirror of the OPO cavity, there is a perfect intensity
correlation between the two modes when they are measured on time
intervals long compared to the cavity storage time \cite{reynaud}.
From this, we infer that the quantum state describing the output
of such a perfect OPO is the eigenvector of
$\hat{I_1}-\hat{I_2}=\hat{a}_1^{+}\hat{a}_1-\hat{a}_2^{+}\hat{a}_2$
with a zero eigenvalue. By assuming a gaussian photon statistics,
the state can be described by
\begin{eqnarray}
|\Psi_{OPO}\rangle=\sum_{n=0}^{+\infty} c_n |n,n\rangle
\end{eqnarray}
with
\begin{eqnarray}
\label{coeff} | c_n |^2 = \frac{1}{\sqrt{2 \pi F \bar n  }} \
\mathrm{exp} \left[ -\frac{(n - \bar n)^2}{2 F \bar n} \right]
\end{eqnarray}
or more generally by a density matrix
\begin{eqnarray}
\rho=\sum_{n,n'} c_{n,n'}|n,n\rangle\langle n',n'|
\end{eqnarray}
with $c_{n,n}$ equal to the expression of $|c_n|^2$ given by
equation (\ref{coeff}). This form of the density matrix ensures at
the same time a Gaussian distribution of photon numbers in each
mode and a perfect intensity correlation between them. It will be
shown later that the unknown expressions of $c_{n,n'}$ for $n\neq
n'$ are not relevant in our calculations.

\begin{figure}
\includegraphics[width=.8\columnwidth]{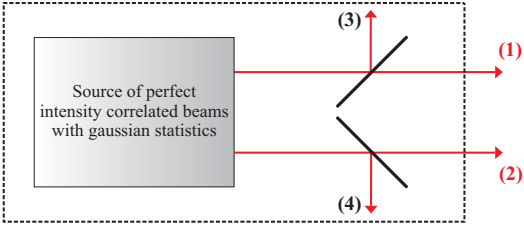}
\caption{\label{figure2}The dotted rectangle models a ND-OPO which
can be described by a source of perfect intensity correlated beams
plus a partition process generated by two symmetric beam-splitters
with intensity transmission $T$ and reflection $R$. The
"gemellity" $G$ of the beams is found to be equal to the loss
coefficient $R$.}
\end{figure}

Uncorrelated losses in the system inevitably degrade the initial
correlation between the signal and idler modes. These linear
losses, that we will assume to be equal for the signal and idler
modes, can be modelled by a beam-splitter of intensity reflection
coefficient $R$ equal to the loss coefficient and of transmission
coefficient $T=1-R$ (Fig. \ref{figure2}). If the input state
impinging on such a beam-splitter is the tensor product
$|n,0\rangle$ of a Fock state on the input mode 1 and the vacuum
in the loss mode, the output state $|\Psi_{out}\rangle$ is
\cite{loudon}
\begin{eqnarray}
|\Psi_{out}\rangle=\sum_{p=0}^{n} A_{n,p} |n-p,p\rangle
\end{eqnarray}
with $A_{n,p}$ given by
\begin{eqnarray}
A_{n,p} =\sqrt{\frac{n!}{p! (n-p)!} T^{n-p} R^p}
\end{eqnarray}
corresponding to a binomial distribution of the photons over the
two outputs of the beam-splitter, characteristic of a partition
process. The system is then described by a pure state belonging to
a four-mode Hilbert space : the signal, idler modes and the two
loss modes of the two outputs. By tracing over these two loss
modes which are not measured, one obtains the following reduced
density matrix
\begin{eqnarray}
\rho'&=&\sum_{n,n',p,p'}c_{n,n'}A_{n,p}A^*_{n',p}A_{n,p'}A^*_{n',p'}
\nonumber \\
 && \qquad \quad |n-p,n-p'\rangle\langle n'-p,n'-p'|
\end{eqnarray}
This expression is valid for any values of the parameters $R, \bar
n$ and $F$. From it, one can calculate the probability
distribution of the intensity difference. Its mean value is zero.
Its variance can be written as
\begin{eqnarray}
\Delta^2(\hat{I_1}-\hat{I_2})&=&\mathrm{Tr}(\ \rho'(\hat{I_1}-\hat{I_2})^{2}\ ) \\
\nonumber &=&\sum_{n=0}^{\infty}|c_{n} |^2\sum_{p,p'=0}^{n}
|A_{n,p}| ^2|A_{n,p'}| ^2(p-p')^{2}
\end{eqnarray}
and becomes
\begin{eqnarray}
\Delta^2(\hat{I_1}-\hat{I_2})=2R\bar n T=2R\bar n' \end{eqnarray}
where $\bar n T=\bar n'$ is the mean photon number of the signal
and idler mode after the beam-splitter.
$\Delta^2(\hat{I_1}-\hat{I_2})$ is independent of $F$ and
coincides with the value obtained by the usual linearized theory
for the fluctuations. From this, one finds that the "gemellity"
$G$, which is the remaining noise on the intensity difference
normalized to the total shot noise level $2\bar n'$, is equal to
the loss coefficient $R$.

As bright beams have very large mean values in comparison with the
Poissonian standard deviation, one can easily assume that both the
reflected and transmitted beams have also large mean values, even
for $R$ or $T$ close to 1. Thus, one can use the Stirling's
approximation and show that, in a normalized form
\begin{eqnarray}
| A_{n,p} |^2 = \frac{1}{\sqrt{2 \pi n R T  }} \ \mathrm{exp}
\left[- \frac{(p-n R)^2}{2 n R T} \right]
\end{eqnarray}

Let us calculate the photon distribution of the individual beams
after the beam-splitters. If we only consider measurements
involving output 1, the reduced density matrix $\rho''$ is
obtained by tracing over the output 2
\begin{eqnarray} \rho'' =
\sum_{n,n',p} c_{n,n'} A_{n,p} A^*_{n',p} |n-p\rangle\langle n'-p|
\end{eqnarray}
The normalized probability distribution for the number of photons
can thus be written as
\begin{eqnarray}
P(n_1) &=& \langle n_1|\rho''|n_1 \rangle
\\\nonumber &=&\sum_{n=0}^{+\infty} |c_{n} |^2
|A_{n,n-n_1}| ^2
\end{eqnarray}
and becomes
\begin{eqnarray}
\quad P(n_1) &=&\sum_{n=0}^{+\infty}
\frac{1}{\sqrt{2 \pi F \bar n  }}\frac{1}{\sqrt{2 \pi n R T  }}\\
&&\nonumber \times\mathrm{exp} \left[- \frac{(nT-\bar n')^2}{2 F
\bar n' T} - \frac{(n-n_1-n R )^2}{2 n R T} \right]
\end{eqnarray}
We introduce $\delta = nT-\bar n'$ and $\varepsilon = n_1-\bar n'
$ and we assume that $n$ can be taken constant and equal to $\bar
n$ in the factor $\frac{1}{\sqrt{2 \pi n R T }}$. As the numbers
of photon are large, the parameters $\delta$ and $\varepsilon$ can
be considered as continuous variables. The sum can thus be
rewritten as an integral
\begin{eqnarray}
P(\varepsilon)= \frac{1}{T} \frac{1}{\sqrt{2 \pi \bar n'
R}}\frac{1}{\sqrt{2 \pi F \bar n
}}\qquad \qquad \qquad \qquad\\
\nonumber  \int_{-\infty}^{+\infty} \mathrm{exp} \left[ -
\frac{\delta^2}{2 F \bar n' T} \right] \mathrm{exp} \left[-
\frac{(\delta  - \varepsilon)^2}{2 R (\delta+\bar n')}\right]
d\delta
\end{eqnarray}
Developed to the first order in $\frac{\delta}{\bar n'}$ and
$\frac{\varepsilon}{\bar n'}$
\begin{eqnarray}
\!\!P(\varepsilon)=\frac{1}{T} \frac{1}{\sqrt{2 \pi F \bar n }}
\frac{1}{\sqrt{2 \pi \bar n' R}} \qquad \qquad \quad\qquad \qquad\\
\nonumber  \int_{-\infty}^{+\infty}\!\! \mathrm{exp} \left[ -
\delta^2 (\frac{1}{2 F \bar n' T}+\frac{1}{2 R \bar n'} ) +\delta
\frac{\varepsilon}{R \bar n'}- \frac{\varepsilon^2}{2 R \bar
n'}\right] d\delta
\end{eqnarray}
and by using the formula
\begin{eqnarray}
\label{transform} \int_{-\infty}^{+\infty} \mathrm{exp} \left[
-A_1 \delta^2+A_2 \delta\right] d\delta = \sqrt{\frac{\pi}{A_1}} \
\mathrm{exp} \left[- \frac{A_2^2}{4 A_1} \right]
\end{eqnarray}
the probability distribution can be integrated to give
\begin{eqnarray}
P(n_1) =  \frac{1}{\sqrt{2\pi\bar n' F'}}\ \mathrm{exp} \left[-
\frac{(n_1-\bar n')^2}{2\bar n' F' } \right]
\end{eqnarray}
with \begin{eqnarray} F'=R+F T
\end{eqnarray}
Thus we find that the new Fano factor $F'$ after the beam-splitter
is given by $R+FT$. In the case of important losses ($R\rightarrow
1$, $T\rightarrow 0$), this Fano factor goes to 1, \emph{i.e.} the
output state distribution is Poissonian as expected.

\section{State reduction by a photon number measurement}
\label{sec:onevalue}In this section, we study the reduced state
resulting from a photon number measurement on beam 2 giving the
pre-assigned value $N$. By tracing over the trigger mode, the
reduced density matrix $\rho'''$ for the signal mode only can be
written as
\begin{eqnarray}
\rho'''=\sum_{n,n',p}
c_{n,n'}A_{n,p}A^*_{n',p}A_{n,n-N}A^*_{n',n-N}\nonumber\\
\qquad\quad|n-p\rangle\langle n'-p|
\end{eqnarray}
One can see that the photon probability distribution depends only
on the diagonal terms of the density matrix $\rho$. The photon
probability distribution of beam 1 when a given value $N$ is
measured on beam 2 is thus
\begin{eqnarray}
P(n_1,n_2=N) = \sum_{n=0}^{+\infty} |c_{n} |^2 |A_{n,n-n_1}| ^2
|A_{n,n-N}| ^2
\end{eqnarray}
With $F'$ and $\bar n'$ the Fano factor and mean value established
in the last section, this distribution can be read as
\begin{eqnarray}
P(n_1,N) &=&\frac{1}{\sqrt{2 \pi F \bar n }} \sum_{n=0}^{+\infty}
\frac{1}{2 \pi n R T}\ \mathrm{exp}
\left[- \frac{(n T-\bar n')^2}{2 (F'-R) \bar n'}  \right. \nonumber\\
&& \left. -\frac{(nT-n_1)^2}{2 n R T} - \frac{(nT-N)^2}{2 n R T}
\right]
\end{eqnarray}
Let us consider $\varepsilon = n_1-\bar n'$ introduced in the
previous section and $\alpha=N-\bar n'$ which corresponds to the
distance between the value of conditioning $N$ and the
distribution center $\bar n'$. Using calculation techniques
similar to the ones of the previous section, the probability
distribution can finally be written as
\begin{eqnarray}
 \label{onevalue}
P(\varepsilon,\alpha) &=& \left(\frac{1}{\sqrt{2\pi \bar n' F'}} \
\mathrm{exp} \left[- \frac{\alpha^2}{2 \bar n' F'}
 \right]\right)\nonumber\\ &&\times\left(\frac{1}{\sqrt{2\pi \bar n' V_c}}\ \mathrm{exp} \left[- \frac{(\varepsilon-\beta \alpha)^2}{2 \bar n' V_c}
 \right]\right)
\end{eqnarray}
 with \begin{eqnarray}
\beta = 1-\frac{G}{F'}
\end{eqnarray}
\begin{eqnarray}
V_c = 2G-\frac{G^2}{F'}
\end{eqnarray}
$V_c$ is the conditional variance of the intensity fluctuations
the signal beam knowing the intensity fluctuations of the trigger
beam. This parameter plays an important role in the
characterization of Quantum Non Demolition measurements
\cite{qnd}.

The first factor between parenthesis in equation (\ref{onevalue})
expresses the preparation probability, i.e. the proportion of
selected points in an experimental implementation. It is maximum
when $\alpha=0$. A value chosen in the wings of the gaussian
distribution leads to a smaller efficiency. The preparation
probability is inversely proportional to $\sqrt{F' \bar n'}$. This
factor is thus very small in the case of bright beams: for  1064
nm beams with a mean power of 1 mW and a Fano factor equal to 100,
the maximal preparation probability obtained when $\alpha=0$ is as
small as $5.10^{-10}$.

The second factor of this probability shows that the selected
state has a gaussian photon distribution centered around the value
$\beta \alpha$, slightly different from the triggering value
$\alpha$ except for $\alpha=0$. This difference can be interpreted
by the dissymmetry of the probability distribution around the
conditioning value. For a conditioning value equal to the mean of
the distribution ($\alpha=0$), this difference vanishes as the
distribution is symmetric.
\begin{figure}
\includegraphics[width=.8\columnwidth]{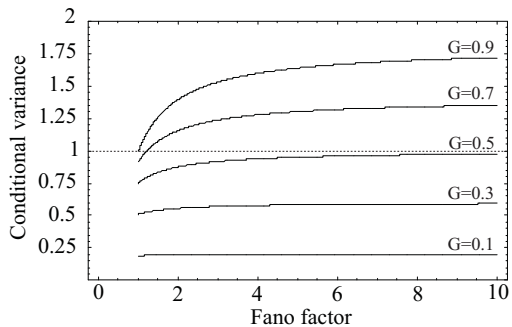}
\caption{\label{figure3}Conditional variance as a function of the
Fano factor for different initial intensity correlation between
signal and idler beams, measured by the "gemellity" $G$. Values of
$G$ as small as $0.1$ have been obtained experimentally
\cite{laurat2}.}
\end{figure}

According to (\ref{onevalue}), the reduced state has a photon
variance equal to the conditional variance $V_c$. This shows that
the selection process we have used in the conditional measurement
has extracted the maximum information available from the
correlation of the two beams and has transferred it to the signal
beam. This beam can exhibit a sub-Poissonian photon distribution
when $V_c<1$. Figure (\ref{figure3}) shows the conditional
variance as a function of the Fano factor for different
gemellities $G$. If the correlation is perfect, this protocol
generates a number state. It is worth noting that for a gemellity
inferior to 0.5 the distribution is sub-Poissonian whatever the
initial Fano factor. For large Fano factors, which is the
experimental case, the noise reduction is twice the initial
gemellity as $V_c$ can be approximated by $2G$.

The conditional variance $V_c$ also characterizes the noise
reduction obtained when an active feed-forward correction of
signal beam intensity is implemented by opto-electronics devices
controlled by information collected on the idler \cite{mertz}. The
two techniques which take advantage of the quantum correlation
have therefore the same ultimate performance.

\section{State reduction by a band selection}
\label{sec:band}As shown in the previous section, triggering on a
single value of photon number in our regime of bright beams where
$\bar n'$ is very large leads to a close to zero preparation
probability incompatible with experimental implementation. An
interesting question is to determine the photon distribution when
one projects onto a finite band instead of a given value. It is
worth pointing out that it is always the experimental case because
of the limited acquisition precision.

Let us consider a conditioning band of width $\Delta$ around the
value $N=\bar n'+\alpha$. The photon probability distribution of
the reduced state can now be written
\begin{eqnarray}
P\left(\varepsilon,\alpha\!-\!\frac{\Delta}{2}\!<\!n_2-\bar n'
\!<\!\alpha\!+\!\frac{\Delta}{2}\!\right) =\frac{1}{\sqrt{2\pi
\bar n' F}
}\frac{1}{\sqrt{2\pi \bar n' V_c}} \nonumber\\
\int^{\alpha+\frac{\Delta}{2}}_{\alpha-\frac{\Delta}{2}}\mathrm{exp}
\left[- \frac{x^2}{2 \bar n' F'}
 \right]\mathrm{exp} \left[- \frac{(\varepsilon-\beta x)^2}{2 \bar n' V_c}
 \right]dx \qquad
\end{eqnarray}
and integrated as
\begin{eqnarray}
\label{exact}
P\left(\varepsilon,\alpha\!-\!\frac{\Delta}{2}\!<\!n_2-\bar
n'\!<\!\alpha\!+\!\frac{\Delta}{2}\!\right)
=\qquad\qquad\qquad\qquad \nonumber\\\frac{1}{\sqrt{2\pi
 F' \bar n'}}\ \mathrm{exp}\left[- \frac{\varepsilon^2}{2 \bar n' F'}
 \right]\qquad\qquad\nonumber\\\times\frac{1}{2}\left(\mathrm{erf}\left[\frac{(\alpha+\frac{\Delta}{2})- \beta \varepsilon}{\sqrt{
2 \bar n'
V_c}}\right]-\mathrm{erf}\left[\frac{(\alpha-\frac{\Delta}{2})-
\beta \varepsilon}{\sqrt{ 2 \bar n' V_c}}\right]\right)\quad
\end{eqnarray}
where erf denotes the error function.
\begin{figure}
\includegraphics[width=8cm]{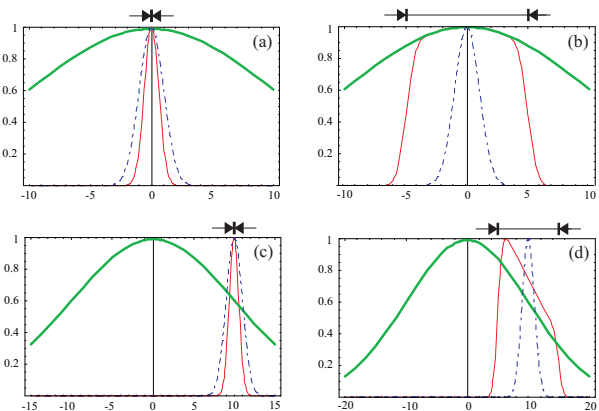}
\caption{\label{figure4}Photon probability distributions of
conditionally prepared state. The probability distributions are
all normalized to 1 at their maximums for clarity of the plots.
The horizontal unit is the width $\sigma=\sqrt{\bar n'}$ of the
Poisson distribution of same mean intensity. Thick curves give the
initial gaussian distribution and dashed ones correspond to a
coherent state of same mean intensity. Arrows give the extension
of the selection band. (a) and (b): the selection band is centered
on the mean value ($\alpha=0$). The bandwidth is
$\Delta=0.1\sigma$ for (a) and $10\sigma$ for (b). A large band
results in a wide distribution (Kurtosis coefficient lower than
3). (c) and (d): the band is centered on $\alpha=10\sigma$.
$\Delta=0.1\sigma$ for (c) and $10\sigma$ for (d). Non central
selection band results in an asymmetry of the distribution (non
null skewness) ($F'=100, G=0.18$)}
\end{figure}

From this exact expression, one can plot the probability
distribution of the conditionally prepared state in different
cases. The initial Fano Factor and gemellity are taken from our
ND-OPO experiment detailed in \cite{laurat} ($F'=100, G=0.18$).
Figure (\ref{figure4}) gives the photon probability distribution
in different possible configurations. The band can be centered
around the mean ($\alpha=0$) or around an arbitrary value taken in
the wings of the initial gaussian distribution ($\alpha\neq0$).
The bandwidth $\Delta$ can also be small or large relative to the
standard deviation $\sigma$ of a Poisson distribution of same mean
intensity. The noise distribution of the initial state and of a
coherent state of same mean intensity have been superimposed. One
observes a narrowing of the probability distribution below the
shot noise level in the case of a very narrow bandwidth for any
band center. When the bandwidth increases, the reduced state
differs more and more from a gaussian distribution.

\begin{figure}[!htpb]
\includegraphics[width=.95\columnwidth]{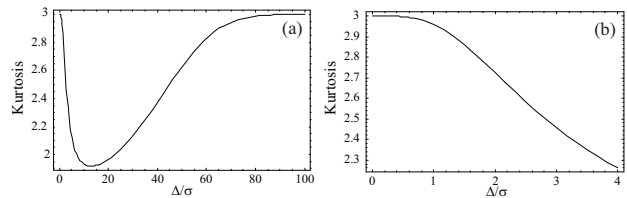}
\caption{\label{figure5}Kurtosis coefficient of the conditionally
selected state according to the selection bandwidth normalized to
$\sigma=\sqrt{\bar n'}$ (b): zoom for small bandwidth. ($\alpha=0,
F'=100, G=0.18$)}
\end{figure}

The distance from a gaussian distribution can be characterized by
the coefficients of skewness and kurtosis which are related to
higher moments of the probability distribution \cite{mandel}. The
coefficient of skewness provides a dimensionless measure of the
asymmetry of the probability distribution. Skewness is null when
the band is centered on the mean value. For a normalized
probability total surface, the kurtosis coefficient distinguishes
distribution which are tall and thin from those that are short and
wide. It has the value of 3 for a gaussian distribution and a
value lower than 3 corresponds to a less peaked distribution. Let
us note that the third order cumulant corresponds to the kurtosis
excess. A very small value of kurtosis is associated to a
square-shaped distribution. A perfect gaussian distribution is
obtained in the limit of very narrow bandwidth -as calculated in
the previous section- or in the case of very large bandwidth which
corresponds to keep all the values and the selected state
corresponds thus to the initial state. A narrow band of selection
results in a close to gaussian distribution with a kurtosis
coefficient almost constant around 3. This coefficient decreases
when the band is extended (Fig. \ref{figure5}).

We can derive from equation (\ref{exact}) an approximate but
analytical expression for the probability distribution by
expanding in powers of $\frac{\Delta}{\sqrt{\bar n'}}$
\begin{eqnarray}
\label{expand}
P\left(\varepsilon,\alpha\!-\!\frac{\Delta}{2}\!<\!n_2-\bar
n'\!<\!\alpha\!+\!\frac{\Delta}{2}\!\right) =
P(\varepsilon,\alpha)\qquad\qquad\quad
\\\nonumber\times\left(\!\Delta + \frac{\Delta^3}{24 \bar n' V_c^{2}}
(-1+\frac{(\alpha - \beta \varepsilon)^2)}{\bar n' V_c} +
o\left(\!\left(\frac{\Delta}{\sqrt{\bar
n'}}\!\right)^{5}\right)\!\right)
\end{eqnarray}
with $P(\varepsilon,\alpha)$ the probability distribution
determined in the last section.

Let us underline that for $\Delta=1$ the photon distribution is
equal to the one established in the last section. This value
corresponds to the minimum bandwidth necessary to discriminate two
consecutive photon-number states.

The third order term in $\Delta$ can be neglected in this
approximation as long as $\Delta \ll \sigma=\sqrt{\bar n'}$. The
preparation probability can then be written
\begin{eqnarray}
\frac{N_{selected}}{N_{total}}=\frac{(\Delta/\sigma)}{\sqrt{2\pi
F'}}\ \mathrm{exp} \left[- \frac{(\alpha/\sigma)^2}{2 F'}
 \right]
\end{eqnarray}
where $N_{total}$ and $N_{selected}$ correspond respectively to
the total number of events and to the number of selected ones by
the conditional protocol.

As a result, conditioning on a narrow band results in a
preparation probability proportional to $\Delta$ without
decreasing the non classical character of the projected state.
This range of $\Delta$ is the optimal one in order to generate a
sub-Poissonian state. With a bandwidth equal to $0.1\sigma$ the
efficiency reaches $0.4\%$.

\begin{figure}
\includegraphics[width=.95\columnwidth]{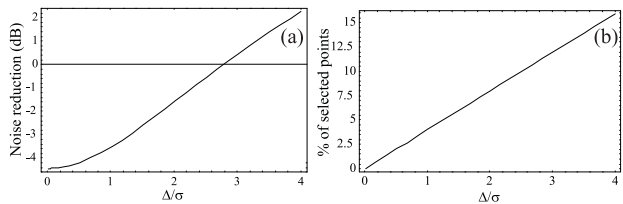}
\caption{\label{figure6} (a) Intensity noise on the reduced state
and (b) preparation probability (proportion of selected points) as
a function of the selection bandwidth normalized to $\sigma$.
($\alpha=0, F'=100, G=0.18$)}
\end{figure}

\begin{figure}[!htpb]
\includegraphics[width=.75\columnwidth]{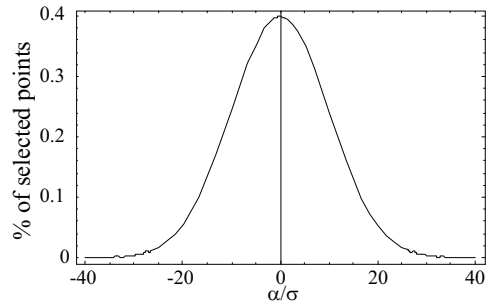}
\caption{\label{figure7} Preparation probability as a function of
the center of the narrow $\Delta=0.1\sigma$ selection band.
($F'=100, G=0.18$)}
\end{figure}

From equation (\ref{exact}), it is possible to compute the exact
properties of the state for any value of $\Delta$. In Fig.
(\ref{figure6}), we give the noise reduction and the probability
of preparation for a narrow bandwidth. One can see that for very
narrow bandwidth the preparation efficiency increases linearly
with the bandwidth whereas the squeezing is almost constant. This
corresponds to the first order in $\Delta$ as appears in
(\ref{expand}). For a larger band, the efficiency still increases
but at the expense of a decreased noise reduction. The efficiency
would be higher for an initial state with less excess noise above
shot noise. This case can correspond to a ND-OPO pumped well above
threshold where it has been shown theoretically that both twin
beams can be individually squeezed \cite{individual} and still
highly correlated. In the case of narrow bandwidth, the band
center has no effect on the non classical character but results in
a lowered preparation efficiency due to the gaussian distribution
of the initial noise (Fig. (\ref{figure7})). It is worth noting
that the large extension of the initial noise should permit to
implement a great number of independent selection bands. Section
\ref{sec:multiband} is devoted to this particular conditional
strategy offered by continuous variable regime.

\section{Monte-Carlo simulation of the state reduction}
\label{sec:simulation}So far we have examined the conditionally
prepared state from its theoretical probability distribution
expression. However, our protocol is easy to test by Monte-Carlo
simulations.

In order to implement the protocol, one needs to prepare two
random arrays $A$ and $B$ with super-Poissonian statistics and
exhibiting a given amount of correlations. These two arrays will
correspond to the fluctuation distribution of the signal and idler
twin beams. Actually, we generate three independent arrays. Let us
call $C$ a super-Poissonian distribution and $X$ and $Y$ two
Poissonian distributions. If the correlation between the signal
and idler beams is perfect, we associate to the signal and idler
beams the same distribution $C$. For a finite amount of
correlations, a similar strategy to the one detailed in section
\ref{sec:opostate} is performed. We add to the initial common
distribution $C$ independent contributions which can be seen as
vacuum contributions when the initial state is incident on a beam
splitter with reflection coefficient $R$. The statistics
distribution of signal and idler after the partition process can
be defined by
\begin{eqnarray}
A=\sqrt{R} ~X+\sqrt{1-R} ~C \nonumber\\
B=\sqrt{R} ~Y+\sqrt{1-R} ~C
\end{eqnarray}
The parameter $R$ is the same as the one used in previous sections
and determines the gemellity of the beams.

The selection of relevant events can then be done. It is worth
noting that conditional measurement experiments are very
frequently made after the end of the physical measurement, i.e. by
post-selection of the events \cite{laurat}. So, when the two
previous distributions have been generated, we dispose of the same
kind of data that after an experiment.

We have simulated all the properties presented before. As
expected, the simulations are in perfect agreement with our
theoretical predictions. Such simulations also should allow to
test the protocol when $\bar n'$ is not too large, a case where
the approximations leading to equation (\ref{onevalue}) are no
longer valid.

\section{Multi-band selection}
\label{sec:multiband}In any conditional protocol, the preparation
probability in a given time interval is an important parameter to
characterize its efficiency. The efficiency of all conditional
preparation techniques is usually very low. Improving the
brightness of the source is thus an important and topical
challenge in the photon counting regime \cite{brightness}. The
preparation probability is also low in our protocol as shown in
Fig. (\ref{figure6}) and (\ref{figure7}) but the continuous
detection results in a great number of selected points in a very
short recording time \cite{laurat}.

\begin{figure}
\includegraphics[width=.95\columnwidth]{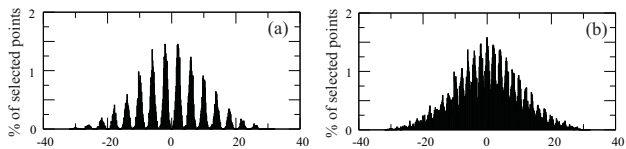}
\caption{\label{figure8} Experimental multi-band selection. The
horizontal unit is the width $\sigma$ of the Poisson distribution
of same mean intensity. Independent bands of width
$\Delta=0.2\sigma$ have been implemented. Band are separated by
$4\sigma$ in (a) and $2\sigma$ in (b). ($F'=100$, $G=0.18$)}
\end{figure}

Furthermore, the continuous variable regime offers the possibility
to improve the efficiency of the conditional measurement strategy
by a large factor. This possibility does not exist for the single
photon counting case. One can implement multiple selection bands
with different centers on the idler intensity. Independent
selection bands will correspond to independent sets of time
windows. In each of them, the state is reduced to a given
sub-Poissonian state with a constant noise reduction, wherever the
band center. By using independent intervals, one keeps most of the
values of the idler intensity and so improve by a large factor the
success rate of the preparation. This scheme -directly related to
the nature of continuous variable- opens the possibility to
generate in parallel a great number of sub-Poissonian states.
Figure (\ref{figure8}) gives the experimental implementation of
multi-band selection for different distances between the band
center. Experimental details are given in \cite{laurat}.

\section{Conclusions}
\label{sec:conclusion}In the photon counting regime, conditional
measurements plays a crucial role as illustrated by the various
schemes theoretically proposed or experimentally implemented. The
extension to the continuous variable regime is of great importance
as the efficiency can be dramatically improved and more complex
triggering schemes implemented.

The required quantum correlation is generated in our protocol by a
ND-OPO above threshold. It has been shown that conditioning on a
given value of the idler intensity leads to the preparation of a
sub-Poissonian state with a noise reduction equal to the
conditional variance, i.e. the initial gemellity minus 3~dB in the
case of a large Fano factor. We have extended this triggering
condition to a finite band. For a band largely narrower than the
standard deviation of a coherent state of same mean intensity, the
preparation probability increases linearly with the bandwidth and
the noise reduction remains almost constant. As opposed to the
discrete variables case, there is a trade-off between the
preparation efficiency and the non classical character of the
selected state. Theoretical calculations and computer simulations
are in very good agreement, and both account very well for the
experimental results detailed in \cite{laurat}.

The next step should be to extend this conditional measurement
strategy to the preparation of more exotic non-classical states as
it is the case in the photon counting regime. One can think for
instance to generate Schr\"{o}dinger-cat-like state. It could be
also of great interest to extend this protocol to EPR beams, in
particular generated by a self-phase-locked OPO where the
anti-correlated phase fluctuations of the twin beams are
accessible \cite{longchambon}.

\begin{acknowledgments}
Laboratoire Kastler-Brossel, of the Ecole Normale Sup\'{e}rieure
and the Universit\'{e} Pierre et Marie Curie, is associated with
the Centre National de la Recherche Scientifique (UMR 8552). A.
Ma\^{\i}tre and T. Coudreau are also at the P\^ole Mat{\'e}riaux
et Ph{\'e}nom{\`e}nes Quantiques FR CNRS 2437, Universit{\'e}
Denis Diderot, 2, Place Jussieu, 75251 Paris cedex 05, France. We
acknowledge support from the European Commission project QUICOV
(IST-1999-13071) and ACI Photonique (Minist\`ere de la Recherche).
\end{acknowledgments}

\end{document}